\documentclass[aps,preprint]{revtex4}%
\usepackage{amsfonts}
\usepackage{amsmath}
\usepackage{amssymb}
\usepackage{graphicx}%
\providecommand{\U}[1]{\protect\rule{.1in}{.1in}}

\begin{document}
\title[ ]{The Classical Aharonov-Bohm Interaction as a Relativity Paradox}
\author{Timothy H. Boyer}
\affiliation{Department of Physics, City College of the City University of New York, New
York, New York 10031, USA}
\keywords{}
\pacs{}

\begin{abstract}
The situation of a charged particle passing down the symmetry axis through a
magnetic toroid presents a relativity paradox; different inertial frames
suggest different forces on the charge and on the toroid due to the
unperturbed systems. \ We review the charge-toroid interaction and suggest
that the magnetic Aharonov-Bohm situation is misunderstood because of
unfamiliarity with the acceleration fields following from the Darwin
Lagrangian, which go unmentioned in recent textbooks of classical
electromagnetism. \ 

\end{abstract}
\maketitle

\section{Introduction}

\subsection{The Aharonov-Bohm Situation}

The magnetic Aharonov-Bohm phase shift involving electrons passing a long
solenoid has attracted great attention because it is claimed to be an effect
of the vector potential involving no forces on the passing electrons and
having no classical analogue.\cite{AB1959}\cite{OP} \ This interpretation of
the observed phenomenon is pervasive in the physics literature.\cite{Feynman}%
\cite{Shadowitz}\cite{Garg} \ In contradiction\cite{BAB} to such views, it is
suggested here that the classical electromagnetic interaction of a charge
particle and a solenoid should be regarded as yet another example of a
relativity paradox where the outcome is easily understood in one interial
frame but is disguised in another. \ 

\subsection{Relativity Paradoxes}

The appearance of relativity paradoxes is familiar to any instructor who has
taught special relativity. \ Perhaps the most famous example is the
pole-and-the-barn paradox where the barn has one open door and a sturdy back
wall.\cite{PandB} \ The \ description in the inertial frame of the barn is
clear. \ The farmer claims that the fast-moving pole is Lorentz contracted and
so easily fits inside the barn before he closes the door. \ The account in the
rest-frame of the pole is misleading, because the physics in this frame
requires new forces which are not mentioned in the original description of the
unperturbed motions of the pole and of the barn. \ Similarly, the
Aharonov-Bohm situation involves two unperturbed systems in relative motion,
in this case a point charge and a solenoid. \ The description is misleading in
the inertial frame where the solenoid is at rest. \ Conservation of energy in
this inertial frame requires forces arising from particle accelerations which
are not mentioned in the original description of the unperturbed motion of the
moving charge and constant-current solenoid. \ 

\subsection{Aharonov-Bohm Situation as Relativity Paradox}

The classical Aharonov-Bohm situation involves the electromagnetic interaction
of a charged particle and a magnet at the relativistic $1/c^{2}$-level, though
this \textit{relativity} aspect is rarely mentioned in the literature. \ The
interaction between the charged particle and the solenoid is calculated in the
approximation that each continues its unperturbed behavior during the
interaction. The interaction is much more easily understood in the interial
frame where the charged particle is at rest and the solenoid is moving,
because in this inertial frame, the physics requires no new forces beyond
those arising from the original descriptions of the unperturbed parts of the
interacting system. \ Indeed, in this inertial frame where the charge is at
rest, it is easy to verify the energy conservation law \ based upon the
equal-and-opposite electric forces that the unperturbed charge and solenoid
put on each other. \ On the other hand, in the inertial frame in which the
solenoid is at rest, energy conservation is violated \textit{unless} one
introduces additional particle accelerations or external forces beyond those
present in the unperturbed solenoid. If the charges of the solenoid are
allowed to accelerate, they introduce back (Faraday) forces on the electron
which were not included in the original description of an \textit{unperturbed}
solenoid. \ Alternatively, one may introduce external forces holding the
solenoid particles at constant speed, and these external forces account for
the required changes in energy, but such external forces were not part of the
original description of the interaction of a charged particle and a solenoid
as \textit{unperturbed} systems.

\subsection{Paradoxes Involving Particle-Magnet Interactions}

The interaction of charges and magnets occurs at the relativistic $1/c^{2}%
$-level of energy and momentum. \ Because the interaction of charges and
magnets at the relativistic level is poorly understood in classical
electrodynamics, it has given rise to a whole class of \textquotedblleft
paradoxes,\textquotedblright\ including the Aharonov-Bohm phase
shift,\cite{AB1959} the Aharonov-Casher phase shift,\cite{AC} the
Shockley-James paradox,\cite{SJ} \textquotedblleft hidden momentum in
magnets,\textquotedblright\cite{hm} and Mansuripur's erroneous
claim.\cite{Mans} \ All of these effects involve relativistic interactions
where our familiar experience with nonrelativistic mechanics, or with
electrostatics, or with magnetostatics may not be adequate. \ These
interactions can all be treated at the level of the Darwin
Lagrangian\cite{Darwin} which describes quasi-static classical electrodynamics
which excludes radiation.

In this article, we will treat only the energy conservation aspects of the
interaction between a point charge and a toroid. \ A more complete description
of the interaction between a point charge and magnet will be published
elsewhere.\cite{B}\ 

\section{Interaction of a Point Charge and a Magnetic Moment\ }

\subsection{Magnetic Dipole Moment}

At its basic level, the problem of the classical Aharonov-Bohm situation
involves the interaction of a magnetic moment $\mathbf{m}$ and a point charge
$e$. \ We will picture the magnetic moment in its own $S_{\mathbf{m}}$ rest
frame as an electrically-neutral circular current loop of radius $b$ and
current $I$. \ The magnetic moment of this current loop (in Gaussian units)
is
\begin{equation}
\mathbf{m}=\mathbf{n}\pi b^{2}I/c,
\end{equation}
where the direction $\mathbf{n}$ is normal to the plane of the current loop
and is connected to the direction of the current $I$ by the right-hand rule.
\ If the center of the current loop is at $\mathbf{r}_{\mathbf{m}}$, we assume
that the point charge $e$ at $\mathbf{r}_{e}$ is sufficiently far away that
the separation is large compared to the radius $b$, $b<<\left\vert
\mathbf{r}_{e}-\mathbf{r}_{\mathbf{m}}\right\vert $, and so the magnetic
dipole approximation is adequate. \ 

\subsection{Interaction in the Inertial Frame where the Magnetic Moment is at
Rest}

In the $S_{\mathbf{m}}$ inertial frame where the magnetic moment is at rest
and the charge $e$ is moving with \textit{constant} velocity $\mathbf{v}%
_{e}=\mathbf{v}$, the charge $e$ carries (through order $1/c^{2}$) both an
electric field
\begin{equation}
\mathbf{E}_{e}\left(  \mathbf{r},t\right)  =e\frac{\left(  \mathbf{r-r}%
_{e}\right)  }{\left\vert \mathbf{r-r}_{e}\right\vert ^{3}}\left[  1+\frac
{1}{2}\frac{v^{2}}{c^{2}}-\frac{3}{2}\left(  \frac{\left(  \mathbf{r-r}%
_{e}\right)  \cdot\mathbf{v}}{\left\vert \mathbf{r-r}_{e}\right\vert
c}\right)  ^{2}\right]  \label{Eev}%
\end{equation}
and a magnetic field
\begin{equation}
\mathbf{B}_{e}\left(  \mathbf{r},t\right)  =e\left(  \frac{\mathbf{v}}%
{c}\right)  \times\frac{\left(  \mathbf{r-r}_{e}\right)  }{\left\vert
\mathbf{r-r}_{e}\right\vert ^{3}},
\end{equation}
so that the charge $e$ has an interaction energy with the magnetic moment
given by the \textit{magnetic} field energy\cite{Griffiths291}%
\begin{equation}
\Delta U^{\left(  B\right)  }=-\mathbf{m\cdot B}_{e}\left(  \mathbf{r}%
_{\mathbf{m}},t\right)  \approx-\mathbf{m\cdot}\left[  e\left(  \frac
{\mathbf{v}}{c}\right)  \times\frac{\left(  \mathbf{r}_{\mathbf{m}}%
\mathbf{-r}_{e}\right)  }{\left\vert \mathbf{r}_{\mathbf{m}}\mathbf{-r}%
_{e}\right\vert ^{3}}\right]  . \label{DUB1}%
\end{equation}
In this inertial frame, the magnetic moment experiences a \textit{magnetic}
force%
\begin{align}
\mathbf{F}_{on\mathbf{m}}^{\left(  B\right)  }  &  =-\nabla_{\mathbf{m}%
}\left[  -\mathbf{m\cdot B}_{e}\left(  \mathbf{r}_{\mathbf{m}},t\right)
\right]  =\nabla_{\mathbf{m}}\left\{  \mathbf{m\cdot}\left[  e\left(
\frac{\mathbf{v}}{c}\right)  \times\frac{\left(  \mathbf{r}_{\mathbf{m}%
}\mathbf{-r}_{e}\right)  }{\left\vert \mathbf{r}_{\mathbf{m}}\mathbf{-r}%
_{e}\right\vert ^{3}}\right]  \right\} \nonumber\\
&  =\nabla_{\mathbf{m}}\left\{  \left[  \mathbf{m\times}e\left(
\frac{\mathbf{v}}{c}\right)  \right]  \cdot\frac{\left(  \mathbf{r}%
_{\mathbf{m}}\mathbf{-r}_{e}\right)  }{\left\vert \mathbf{r}_{\mathbf{m}%
}\mathbf{-r}_{e}\right\vert ^{3}}\right\} \nonumber\\
&  =\left[  \mathbf{m\times}e\left(  \frac{\mathbf{v}}{c}\right)  \right]
\cdot\nabla_{\mathbf{m}}\left(  \frac{\left(  \mathbf{r}_{\mathbf{m}%
}\mathbf{-r}_{e}\right)  }{\left\vert \mathbf{r}_{\mathbf{m}}\mathbf{-r}%
_{e}\right\vert ^{3}}\right) \nonumber\\
&  =\frac{-3\left[  \mathbf{m\times}e\left(  \mathbf{v}/c\right)  \right]
\cdot\left(  \mathbf{r}_{\mathbf{m}}\mathbf{-r}_{e}\right)  \left(
\mathbf{r}_{\mathbf{m}}\mathbf{-r}_{e}\right)  }{\left\vert \mathbf{r}%
_{\mathbf{m}}\mathbf{-r}_{e}\right\vert ^{5}}+\frac{\left[  \mathbf{m\times
}e\left(  \mathbf{v}/c\right)  \right]  }{\left\vert \mathbf{r}_{\mathbf{m}%
}\mathbf{-r}_{e}\right\vert ^{3}}, \label{FBonm}%
\end{align}
while the charge $e$ experiences a (deflecting) \textit{magnetic} force due to
the magnetic dipole
\begin{equation}
\mathbf{F}_{on\mathbf{e}}^{\left(  B\right)  }=e\frac{\mathbf{v}}{c}%
\times\mathbf{B}_{\mathbf{m}}\left(  \mathbf{r}_{e},t\right)  =e\frac
{\mathbf{v}}{c}\times\left[  \frac{3\mathbf{m\cdot}\left(  \mathbf{r}%
_{e}\mathbf{-r}_{\mathbf{m}}\right)  \left(  \mathbf{r}_{e}\mathbf{-r}%
_{\mathbf{m}}\right)  }{\left\vert \mathbf{r}_{e}\mathbf{-r}_{\mathbf{m}%
}\right\vert ^{5}}-\frac{\mathbf{m}}{\left\vert \mathbf{r}_{e}\mathbf{-r}%
_{\mathbf{m}}\right\vert ^{3}}\right]  . \label{FBone}%
\end{equation}
In this inertial frame, the forces between the magnetic moment and the charge
are not equal in magnitude and opposite in direction; Eq. (\ref{FBonm})
involves a term in the direction $\left(  \mathbf{r}_{\mathbf{m}}%
\mathbf{-r}_{e}\right)  $ whereas Eq. (\ref{FBone}) involves a term in the
dirction $\mathbf{v\times}$ $\left(  \mathbf{r}_{e}\mathbf{-r}_{\mathbf{m}%
}\right)  $. \ 

\subsection{Interaction in the Inertial Frame where the Charge $e$ is at Rest}

On the other hand, in the $S_{e}$ inertial frame where the charge particle $e$
is at rest and the magnetic moment is moving with velocity $\mathbf{v}%
_{\mathbf{m}}=-\mathbf{v}$, the interaction between the charge and the
magnetic moment involves energy in the \textit{electric} fields because, in
this frame where it is moving, the magnetic moment has an \textit{electric}
dipole moment\cite{Jackson1}
\begin{equation}
\mathbf{p}_{\mathbf{m}}\approxeq\left(  \frac{-\mathbf{v}}{c}\right)
\times\mathbf{m.} \label{pmv}%
\end{equation}
In this $S_{e}$ inertial frame, the electric interaction energy is\cite{G172}
\begin{equation}
\Delta U^{\left(  E\right)  }=-\mathbf{p}_{\mathbf{m}}\cdot\mathbf{E}%
_{e}\left(  \mathbf{r}_{\mathbf{m}},t\right)  =-\left[  \frac{-\mathbf{v}}%
{c}\times\mathbf{m}\right]  \cdot\left(  e\frac{\left(  \mathbf{r}%
_{\mathbf{m}}\mathbf{-r}_{e}\right)  }{\left\vert \mathbf{r}_{\mathbf{m}%
}\mathbf{-r}_{e}\right\vert ^{3}}\right)  ,
\end{equation}
which is the same as the \textit{magnetic} energy given in Eq. (\ref{DUB1}).
\ The \textit{electric} force on the magnetic moment is accordingly
\begin{align}
\mathbf{F}_{on\mathbf{m}}^{\left(  E\right)  }  &  =-\nabla_{\mathbf{m}%
}\left[  -\mathbf{p}_{\mathbf{m}}\cdot\mathbf{E}_{e}\left(  \mathbf{r}%
_{\mathbf{m}},t\right)  \right]  =\left(  \mathbf{p}_{\mathbf{m}}\cdot
\nabla_{\mathbf{m}}\right)  \mathbf{E}_{e}\left(  \mathbf{r}_{\mathbf{m}%
},t\right) \nonumber\\
&  =\left\{  \left[  \frac{-\mathbf{v}}{c}\times\mathbf{m}\right]  \cdot
\nabla_{\mathbf{m}}\right\}  \left(  e\frac{\left(  \mathbf{r}_{\mathbf{m}%
}\mathbf{-r}_{e}\right)  }{\left\vert \mathbf{r}_{\mathbf{m}}\mathbf{-r}%
_{e}\right\vert ^{3}}\right) \nonumber\\
&  =\frac{-3\left[  \left(  -\mathbf{v/c}\right)  \times\mathbf{m}\right]
\cdot\left(  \mathbf{r}_{e}\mathbf{-r}_{\mathbf{m}}\right)  \left(
\mathbf{r}_{e}\mathbf{-r}_{\mathbf{m}}\right)  }{\left\vert \mathbf{r}%
_{\mathbf{m}}\mathbf{-r}_{e}\right\vert ^{5}}+\frac{\left[  \left(
-\mathbf{v/c}\right)  \times\mathbf{m}\right]  }{\left\vert \mathbf{r}%
_{\mathbf{m}}\mathbf{-r}_{e}\right\vert ^{3}}. \label{FEonm}%
\end{align}
Noting the reversals of sign connected with the order in the cross products,
one finds that this electric force on the magnetic dipole in Eq. (\ref{FEonm})
is the same as the magnetic force on the magnetic dipole appearing in Eq.
(\ref{FBonm}). \ Also, the \textit{electric} force on the charge $e$ is just
the negative of this expression,%
\begin{equation}
\mathbf{F}_{on\mathbf{e}}^{\left(  E\right)  }=e\mathbf{E}_{\mathbf{m}}\left(
\mathbf{r}_{e},t\right)  =e\left\{  \frac{3\left[  \left(  -\mathbf{v/c}%
\right)  \times\mathbf{m}\right]  \cdot\left(  \mathbf{r}_{e}\mathbf{-r}%
_{\mathbf{m}}\right)  \left(  \mathbf{r}_{e}\mathbf{-r}_{\mathbf{m}}\right)
}{\left\vert \mathbf{r}_{e}\mathbf{-r}_{\mathbf{m}}\right\vert ^{5}}%
-\frac{\left[  \left(  -\mathbf{v/c}\right)  \times\mathbf{m}\right]
}{\left\vert \mathbf{r}_{e}\mathbf{-r}_{\mathbf{m}}\right\vert ^{3}}\right\}
. \label{FEone}%
\end{equation}
Through order $1/c^{2}$ in this $S_{e}$ inertial frame, the electric forces
that the magnetic moment and charge place on each other are equal in magnitude
and opposite in direction. \ The change in electric field energy is accounted
for by the work done by the electric force $\mathbf{F}_{on\mathbf{m}}^{\left(
E\right)  }$ on the moving magnetic moment. \ 

\section{Transition to a Point Charge and Toroid}

\subsection{Forming a Toroid from Magnetic Dipoles}

Although the equations which we have listed already record the basic paradox,
the situation becomes far more vivid, and also simpler calculationally, if we
imagine \ many magnetic moments arranged so as to form a toroid. \ And indeed,
a toroid can be pictured as a solenoid (a stack of current loops) which is
bent into a circular shape and so brings us to the Aharonov-Bohm situation
where electrons pass a long solenoid. \ 

Thus, we picture the magnetic moments (which are simply circular current loops
of radius $b$) arranged in a circular pattern of (average) radius $R$ \ around
the $z$-axis so as to form a toroid located along the $z$-axis at $z_{T}$.
\ Each current loop lies in the plane formed by the $z$-axis and the
displacement from the $z$-axis to the center of the current loop. \ We assume
that there are $N$ current loops, each carrying current $I$ and that the
(average) radius $R$ of the toroid is much larger than the radius $b$ of each
current loop, $b<<R$. \ The average magnetic field inside the toroid is
\begin{equation}
\mathbf{B}_{T}=\widehat{\phi}\frac{4\pi}{c}\frac{NI}{2\pi R}=\widehat{\phi
}\frac{2NI}{cR},
\end{equation}
and the magnetic flux through each current loop of the toroid is
\begin{equation}
\Phi=\pi b^{2}B_{T}=\frac{2\pi b^{2}NI}{cR}.
\end{equation}
For the electrically-neutral toroidal situation, there are no toroidal
electric fields, and all the magnetic fields are confined to the interior of
the toroid.

\subsection{Interaction of a Charge and a Toroid}

We consider a charged particle $e$ moving with velocity $\mathbf{v}%
_{e}=\widehat{z}v$ along the $z$-axis, which is the axis of symmetry of the
toroid. \ We want to obtain the lowest non-vanishing approximation for the
interaction between the charge $e$ and the toroid. \ This \textquotedblleft
lowest-nonvanishing-interaction\textquotedblright\ approximation suggests that
we consider the toroid and the charge $e$ as continuing their
\textit{unperturbed} motions despite their mutual interaction. Thus we
consider the currents carried by the charge carriers of the toroid as
constant. \ We also consider the velocity $\mathbf{v}_{e}$ of the charge $e$
as constant. \ With these assumptions, we wish to determine the forces on the
charge $e$ and on the toroid due to the toroid and the charge $e$ respectively
through order $1/c^{2}$.

\subsection{Toroid at Rest}

In the $S_{T}$ inertial frame where the toroid is at rest and the charge $e$
is moving with velocity $\mathbf{v}_{e}\mathbf{=}\widehat{z}v$, it appears
that the passing charge puts a magnetic force (corresponding to Eq.
(\ref{FBonm})) on each magnetic dipole moment (circular current loop) of the
toroid. \ By symmetry, the $z$-components of the forces add while the radial
components cancel. \ The magnetic field of the charge $e$ (assumed positive)
is in the same circular pattern as that of the toroid. \ Taking the negative
derivative of the $-\mathbf{m\cdot B}$ contributions gives a total magnetic
force on the toroid
\begin{align}
\mathbf{F}_{onT}^{\left(  B\right)  }  &  =-\widehat{z}\frac{\partial
}{\partial z_{T}}\left(  -N\frac{\pi b^{2}I}{c}B_{e}(z_{T},t\right)
\nonumber\\
&  =\widehat{z}\frac{\partial}{\partial z_{T}}\left(  \left(  \frac{N\pi
b^{2}I}{c}\right)  e\frac{v}{c}\frac{R}{\left\vert \left(  z_{T}-z_{e}\right)
^{2}+R^{2}\right\vert ^{3/2}}\right) \nonumber\\
&  =\widehat{z}\left(  \frac{N\pi b^{2}I}{c}\right)  e\frac{v}{c}%
\frac{R\left[  -3\left(  z_{T}-z_{e}\right)  \right]  }{\left\vert \left(
z_{T}-z_{e}\right)  ^{2}+R^{2}\right\vert ^{5/2}}. \label{FTB2}%
\end{align}

Since the toroid is electrically neutral and all the magnetic fields of the
toroid are confined to the interior of the toroid, there appears to be no
force back of the \textit{unperturbed} toroid on the charge $e$. \ 

Since the charge particle $e$ experiences no forces, there is no change in its
kinetic energy. \ Since the toroid is electrically neutral, there is no change
in the electric energy as the charge $e$ and the toroid interact. \ However,
there is a change in the system magnetic energy associated with the overlap of
the magnetic field of the charge with the magnetic field of the toroid in the
volume of the toroid,%
\begin{equation}
\Delta U_{overlap}^{\left(  B\right)  }=\frac{1}{4\pi}%
{\textstyle\int}
d^{3}r\mathbf{B}_{e}\cdot\mathbf{B}_{T}=\frac{1}{4\pi}\left(  \frac
{evR}{c\left[  \left(  z_{T}-z_{e}\right)  ^{2}+R^{2}\right]  ^{3/2}}\right)
\left(  \frac{2NI}{cR}\right)  \left[  2\pi R\pi b^{2}\right]  . \label{DUB}%
\end{equation}
\ In this inertial frame, it may appear that the relativistic conservation law
of energy is violated, since there is apparently no force on the moving charge
$e$ and the currents of the toroid are assumed unperturbed.

\subsection{Charge $e$ at Rest}

On the other hand, in the $S_{e}$ inertial frame in which the charge $e$ is at
rest while the unperturbed toroid is moving with velocity $\mathbf{v}%
_{T}=-\widehat{z}v\,$, there are \textit{electric} forces between the charged
particle and the toroid. \ In an inertial frame in which it is moving with
velocity $-\mathbf{v}$, an unperturbed magnetic moment $\mathbf{m}$ has an
\textit{electric} dipole moment $\mathbf{p}_{\mathbf{m}}=\left(
-\mathbf{v/}c\right)  \times\mathbf{m}$ as given in Eq. (\ref{pmv}). \ Thus in
the $S_{e}$ inertial frame in which it is moving with velocity $-\mathbf{v}$,
the \textit{unperturbed} toroid has a ring of \textit{electric} dipoles which
produce a net $z$-component of electric force on the charge $e$ which is $N$
times larger than the $z$-component of force produced by a single electric
dipole in Eq. (\ref{FEone}),%
\begin{equation}
\mathbf{F}_{on\mathbf{e}}^{\left(  E\right)  }=e\mathbf{E}_{T}\left(
\mathbf{r}_{e},t\right)  =\widehat{z}e\frac{N3R\left(  z_{T}-z_{e}\right)
}{c\left[  \left(  z_{T}-z_{e}\right)  ^{2}+R^{2}\right]  ^{5/2}}\left(
\frac{v\pi b^{2}I}{c}\right)  .
\end{equation}
Also, the charge $e$ will place an \textit{electric} force on each electric
dipole of the moving toroid, giving a net force on the toroid%
\begin{align}
\mathbf{F}_{onT}^{\left(  E\right)  }  &  =\widehat{z}\left\{  N\widehat{z}%
\cdot\left[  \left(  \mathbf{p}_{\mathbf{m}}\cdot\nabla_{\mathbf{m}}\right)
\mathbf{E}_{e}\left(  \mathbf{r}_{\mathbf{m}},t\right)  \right]  \right\}
\nonumber\\
&  =\widehat{z}N\widehat{z}\cdot\left[  \left(  p_{\mathbf{m}}\frac{\partial
}{\partial r}\right)  e\frac{\widehat{r}r+\widehat{z}\left(  z_{T}%
-z_{e}\right)  }{\left[  \left(  z_{T}-z_{e}\right)  ^{2}+r^{2}\right]
^{3/2}}\right]  _{r=R}\nonumber\\
&  =\widehat{z}N\left[  p_{\mathbf{m}}e\frac{-3r\left(  z_{T}-z_{e}\right)
}{\left[  \left(  z_{T}-z_{e}\right)  ^{2}+r^{2}\right]  ^{5/2}}\right]
_{r=R}\nonumber\\
&  =\widehat{z}N\left[  \left(  \frac{v}{c}\frac{\pi b^{2}I}{c}\right)
e\frac{-3R\left(  z_{T}-z_{e}\right)  }{\left[  \left(  z_{T}-z_{e}\right)
^{2}+R^{2}\right]  ^{5/2}}\right]  . \label{FTE2}%
\end{align}
This \textit{electric} force on the toroid in Eq. (\ref{FTE2}) is exactly the
same as the \textit{magnetic} force as found in Eq. (\ref{FTB2}) for the
previous inertial frame where the toroid was at rest and the charge $e$ was
moving. \ However, here in the $S_{e}$ inertial frame where the charge $e$ is
at rest and the toroid is moving, the electric forces on the charge $e$ and on
the toroid are equal in magnitude and and opposite in direction.

During the interaction, there is no energy change in the magnetic field
energy, since the charge $e$ is at rest and so has no magnetic field.
\ However, during the interaction, there is a change in the electric field
energy given by
\begin{align}
\Delta U^{\left(  E\right)  }  &  =-N\mathbf{p}_{\mathbf{m}}\cdot
\mathbf{E}_{e}\left(  \mathbf{r}_{\mathbf{m}},t\right) \nonumber\\
&  =-N\left(  \frac{v}{c}\frac{\pi b^{2}I}{c}\right)  E_{er}\left(
\mathbf{r}_{\mathbf{m}},t\right) \nonumber\\
&  =-N\left(  \frac{v}{c}\frac{\pi b^{2}I}{c}\right)  e\frac{R}{\left[
\left(  z_{T}-z_{e}\right)  ^{2}+r^{2}\right]  ^{3/2}}. \label{DUE}%
\end{align}
The electric energy change $\Delta U^{\left(  E\right)  }$ in Eq. (\ref{DUE})
is accounted for by the electric force $\mathbf{F}_{onT}^{\left(  E\right)  }$
on the moving toroid,
\begin{equation}
\Delta U^{\left(  E\right)  }=-%
{\textstyle\int\nolimits_{\infty}^{z_{\mathbf{m}}}}
\mathbf{F}_{T}^{\left(  E\right)  }\cdot\widehat{z}dz_{T}.
\end{equation}
Thus energy conservation involving the unperturbed parts of the system indeed
holds in the $S_{e}$ inertial frame in which the toroid is moving with
velocity $-\mathbf{v}$ and the charge $e$ is at rest. \ 

On the other hand, the change in \textit{electric} energy $\Delta U^{\left(
E\right)  }$ in Eq. (\ref{DUE}) is exactly equal to the \textit{negative} of
the change in \textit{magnetic} energy $\Delta U_{overlap}^{\left(  B\right)
}$ in Eq. (\ref{DUB}) due to the overlap of the magnetic field of the charge
$e$ and the magnetic field of the toroid. \ 

\section{Discussion of the Relativity Paradox}

\subsection{Contrast in Forces Between Different Inertial Frames}

Thus we have our relativity \textquotedblleft paradox.\textquotedblright\ \ In
both inertial frames, all the forces are of order $1/c^{2}$ and so the forces
cannot change in leading order in $v/c$ when viewed from a different inertial
frame. \ Nevertheless, different inertial frames claim that different forces
appear. \ When described in the $S_{T}$ rest frame of the toroid, there is a
\textit{magnetic} force on the toroid, but apparently no force on the moving
charge $e$. \ However, when described in the $S_{e}$\ rest frame of the charge
$e$, there are \textit{electric} forces on the charge $e$ and also on the
toroid. \ Indeed, in the rest frame of the charge $e,$ one finds exactly the
same force on the toroid (now an \textit{electric} force) as was found as a
\textit{magnetic} force in the inertial frame where the toroid is at rest, but
now one also finds its partner in the \textit{electric} force of the toroid on
the charge $e$. \ 

\subsection{The Inertial Frame with the Unreliable Description}

Just as in the relativity paradox of the pole and the barn, one must make a
choice. \ Which description should one trust as representing accurate physics?
\ We suggest that in each case, the accurate description involves the inertial
frame in which the physics does not require the introduction of external
forces and/or accelerations which were not part of the original account of the
\textit{unperturbed} motion. \ For the pole and the barn, the unreliable
description involves the inertial frame in which the barn is moving, and so is
Lorentz contracted; this inertial frame requires the introduction of new
forces when the front of the pole encounters the back wall of the barn, before
the barn door is closed.\cite{PandB} \ These external forces alter the account
given for the \textit{unperturbed} motion of the pole. \ 

In the situation of the classical Aharonov-Bohm interaction of a charged
particle and a magnet, the situation involves the same basic idea. \ In which
inertial frame does the physics require the introduction of new forces and/or
accelerations which were not part of the original account of unperturbed
motion? \ The answer is that the $S_{T}$\ inertial frame in which the toroid
is at rest is less satisfactory; specifically, the changes in magnetic energy
associated with the overlap of the magnetic field of the charge $e$ and the
magnetic field of the toroid have not been accounted for satisfactorily.

\subsection{Problems Involving Magnetic Energy Changes}

Indeed, changes in magnetic energy often present problems. They are the basis
of the present paradox. \ \textit{Electric} energy changes involve work done
directly by the \textit{electric} forces, as is evident in the second
description given for our charge-magnetic interaction where the charge $e$ is
at rest and the toroid is moving. \ In contrast, \textit{magnetic} forces do
no work. \ Therefore \textit{magnetic} energy changes require work being done
by separate \textit{electric or external} forces. \ \textit{Magnetic energy
balance in quasistatic systems requires the existence of electric forces
associated with the accelerations involving changing speeds of charge
particles.} \ Such accelerations are not contained in the description of the
\textit{unperturbed} toroid.

\subsection{Balancing Magnetic Energy Changes for the Toroid at Rest}

The energy balance for the system of the charge $e$ and the toroid involves
three different contributions, mechanical kinetic energy, electric energy, and
magnetic energy%
\[
\Delta U=\Delta U^{\left(  M\right)  }+\Delta U^{\left(  E\right)  }+\Delta
U^{\left(  B\right)  }.
\]
The troublesome aspect, as usual, involves the magnetic energy $\Delta
U^{\left(  B\right)  }$. \ Although the $1/c^{2}$-force on the toroid (given
in Eqs. (\ref{FTB2}) and (\ref{FTE2})) is exactly the same in either inertial
frame, the $1/c^{2}$-energy change of the system given in Eqs. (\ref{DUB}) and
(\ref{DUE}) is \textit{not} the same, but indeed involves a relative
\textit{minus} sign. \ The difficulty here involves the same aspect which
appears in any discussion of magnetic energy changes for quasistatic
systems.\cite{acc}\cite{flat} \ There is a sharp contrast between electric and
magnetic energy changes. \ Electric energies involve only the \textit{relative
positions} between charged particles. \ However, quasistatic magnetic energies
involve \textit{moving} charges. \ Therefore magnetic energy changes can
involve changes in 1) the \textit{relative positions} of the current carriers
and/or in 2) the \textit{speeds} of the charge carriers. \ 

For our charge-toroid example in the $S_{T}$ inertial frame in which the
toroid is at rest and the charge $e$ is moving, we have both aspects of
magnetic energy change,
\[
\Delta U^{\left(  B\right)  }=\Delta U_{overlap}^{\left(  B\right)  }+\Delta
U_{toroid\text{ }currents}^{\left(  B\right)  }.
\]
There is a positive magnetic energy change $\Delta U_{overlap}^{\left(
B\right)  }$ \ associated with the overlap of the magnetic field of the charge
$e$ with the magnetic field of the toroid. However, the electric fields of the
charge $e$ act on the current carriers of the toroid. The zero--order
electrostatic field of the charge $e$ has no $emf$ and so does not deliver net
energy to the toroid currents. \ It is the terms of order $v^{2}/c^{2}$ in Eq.
(\ref{Eev}) which do indeed produce an $emf$ and deliver net energy to (or
remove magnetic energy from) the toroid currents . \ The toroid responds to
the effort to change the speeds of the current carriers\cite{B2015a} in the
fashion typical of a solenoid. \ The (small) accelerations of the (many)
toroid current carriers produce a back (Faraday) acceleration electric field
acting on the agent causing the original $emf$, in this case on the charge
$e$. \ The magnetic energy change due to the changing toroid currents involves
$\mathbf{B}_{T}^{2}$ and so is twice as large and of opposite sign as the
overlap magnetic energy change which involves only the first power of
$\mathbf{B}_{T}$. \ \ It is the back (Faraday) acceleration electric field of
these accelerating charge carriers which places a force on the charge $e$ in
the $S_{T}$ inertial frame where the toroid is at rest and the charge $e$ is
moving. \ The energy-balancing back force is or order $1/c^{2}$. \ 

The electric force on the charge $e$ appears immediately in the
\textit{unperturbed}-motion discussion in the $S_{e}$ inertial frame in which
the toroid is moving and so (according to the \textit{relativistic}
description of the \textit{unperturbed} toroid motion) has an electric dipole
moment. \ In the $S_{T}$ rest frame of the toroid, the basis for the back
field on the charge $e$ involves particle accelerations which are not part of
the description of the \textit{unperturbed} toroid. \ Thus the
\textit{unperturbed} description in the $S_{T}$\ restframe of the toroid,
which does not mention the fields arising from the accelerations of the
current carriers, is indeed the less reliable description of the relativity paradox.

\subsection{Absence of Quasistatic Acceleration Terms in Recent Textbooks}

The back (Faraday) acceleration fields (which are unfamiliar in the
interaction of a charge $e$ and a toroid) are thoroughly familiar in the case
of a solenoid with increasing currents. \ The back $emf$ appearing in a
solenoid when the currents are increasing is caused by these same back
(Faraday) acceleration fields of the accelerating current carriers of the
solenoid.\cite{B2015a} \ However, in the current textbooks of classical
electromagnetism, the solenoid's back $emf$ is calculated from a changing
magnetic flux for a highly-symmetric solenoid, not from the accelerations of
the current carriers. \ 

Acceleration electric fields appear immediately from the Darwin Lagrangian.
\ Thus, at the quasistatic $1/c^{2}$-level, the electric field of an
\textit{accelerating} charge $e$ is \textit{not} that given in Eq. (\ref{Eev})
for a \textit{constant-velocity} charge $e$, but rather includes additional
acceleration-dependent terms,\cite{older}\cite{add}%
\begin{align}
\mathbf{E}\left(  \mathbf{r,}t\right)   &  =%
{\textstyle\sum\nolimits_{a}}
\frac{q_{a}\left(  \mathbf{r}-\mathbf{r}_{a}\right)  }{\left\vert
\mathbf{r}-\mathbf{r}_{a}\right\vert ^{3}}\left[  1+\frac{1}{2}\left(
\frac{\mathbf{\dot{r}}_{a}}{c}\right)  ^{2}-\frac{3}{2}\left(  \frac
{\mathbf{\dot{r}}_{a}\cdot\left(  \mathbf{r}-\mathbf{r}_{a}\right)
}{\left\vert \mathbf{r}-\mathbf{r}_{a}\right\vert c}\right)  ^{2}\right]
\nonumber\\
&  -%
{\textstyle\sum\nolimits_{a}}
\frac{q_{a}}{2c^{2}}\left[  \frac{\mathbf{\ddot{r}}_{a}}{\left\vert
\mathbf{r}-\mathbf{r}_{a}\right\vert }+\frac{\left(  \mathbf{r}-\mathbf{r}%
_{a}\right)  \left[  \mathbf{\ddot{r}}_{a}\cdot\left(  \mathbf{r}%
-\mathbf{r}_{a}\right)  \right]  }{\left\vert \mathbf{r}-\mathbf{r}%
_{a}\right\vert ^{3}}\right]  . \label{Eacc}%
\end{align}
However, even as the Darwin Lagrangian, is barely mentioned in the recent
textbooks of classical electromagnetism, the local (Faraday) acceleration
fields in Eq. (\ref{Eacc}) for an accelerating charge are never mentioned.
\ Fields due to accelerating charges appear only in the sections on radiation
leading to Larmor's formula. \ 

\subsection{Classical Counterpart to the Aharonov-Bohm Effect}

The interaction between a charged particle and a magnet is a relativistic
effect of order $1/c^{2}$ to lowest order. \ Therefore the interaction is
adequately described by the Darwin Lagrangian which reproduces classical
electrodynamics through order $1/c^{2}$ but excludes radiation. \ We expect
that the same basic interaction continues to hold for full classical
electrodynamics, where we have the additional complications of retarded times
and (very small) radiation effects.

It seems widely accepted that there is \textquotedblleft no classical analogue
to the Aharonov-Bohm effect.\textquotedblright\ \ Statements of this sort
appear in many textbooks of quantum theory\cite{quantum} and in some textbooks
of classical electromagnetism.\cite{Shadowitz}\cite{Garg} \ The usual argument
for this no-classical-analogue statement notes that the magnetic field
vanishes outside a very long solenoid or toroid where the currents are
constant, and hence concludes that there is no force on a passing charged
particle. \ However, such unsophisticated views based upon magnetostatics do
not do justice to the subtleties of classical electrodynamics. \ Because
physicists are unfamiliar with the idea of quasistatic accelerating charges
producing the electric fields associated with an $emf$, the claims associated
with the classical Aharonov-Bohm situation have been rarely
challenged.\cite{full}

\section{Acknowledgement}

The reanalysis here of the classical interaction of a charged particle and a
magnet was stimulated by a manuscript of Dr. Hanno Ess\'{e}n,
\textquotedblleft A classical Aharonov-Bohm effet arises when one goes beyond
the test particle approximation.\textquotedblright\ \ I wish to thank Dr.
Ess\'{e}n for alerting me to the work included in reference 21,

\bigskip February 3 AB-asParadox4.tex


\begin{thebibliography}{99}
\bibitem {full}A full discussion of the forces and energy changes is given in
reference \cite{B}.
\end{thebibliography}
\end{document}